# The value of travel speed

Author: Kees van Goeverden

Affiliation: Delft University of Technology

**Abstract**

Travel speed is an intrinsic feature of transport, and enlarging the speed is considered as beneficial. The benefit of a speed increase is generally assessed as the value of the saved travel time. However, this approach conflicts with the observation that time spent on travelling is rather constant and might not be affected by speed changes. The paper aims to define the benefits of a speed increase and addresses two research questions. First, how will a speed increase in person transport work out, which factors are affected? Second, is the value of time a good proxy for the value of speed? Based on studies on time spending and research on the association between speed and land use, we argue that human wealth could be the main affected factor by speed changes, rather than time or access. Then the value of time is not a good proxy for the value of speed: the benefits of a wealth increase are negatively correlated with prosperity following the law of diminishing marginal utility, while the calculated benefits of saved travel time prove to be positively correlated. The inadequacy of the value of time is explained by some shortcomings with respect to the willingness to pay that is generally used for assessing the value of time: people do not predict correctly the personal benefits that will be gained from a decision, and they neglect the social impacts.

**Keywords**: travel speed, travel time, access, human wealth, assessment, happiness.

1. Introduction

Travel speed is an intrinsic feature of transport. If there is no speed, there is no transport. The opportunity to move to another location or to transport goods is essential for human living; the human nature is not equipped for permanently staying at one location, like plants and trees. This means that the value of being able to move is extremely high. The movements will necessarily have a certain speed. Then the question arises whether the magnitude of the speed matters: is moving with speed A more (or less) beneficial than moving with the lower speed B? In practice, we observe that the speed of travelling varies largely among trips, while the average speed tends to increase over time. The paper discusses the value of speed, which we define as the benefit of a marginal speed increase. Will implementations in the transport system that increase travel speed generate benefits that stem from the higher speed? The discussion is limited to the 'internal' benefits; the (generally negative) benefits connected with the external impacts of a speed increase, like the impacts on traffic safety, energy consumption/pollution, and noise nuisance, are left out of consideration. The focus of the discussion is on person travel.

We are not aware of any study that directly addresses the value of speed. Travel speed has no clear benefit in itself, except for some special cases like people who get a kick of travelling at high speed. Possible benefits are related to how speed changes work out. Speed is defined by two variables: distance and time. A speed increase implies either a reduction in travel time for travelling a certain distance, or an increase of the distance travelled in a certain time period. The benefits stem from the reduced travel time or increased travel distance.

The assessment of the benefits of speed changes traditionally focuses on travel time. Research on the value of travel time has a long history and produced a wealth of literature (summarized by Gunn, 2007; Jara-Díaz, 2007; Ortúzar and Willumsen, 2011). The general assumption is that travel time changes are the only or by far most important outcomes of speed changes, implying that people will reallocate their time spending. Gálvez and Jara-Díaz (1998) state that the relevance of the valuation of travel time savings "comes from the obvious fact of travel time reductions being the main source of benefits in transport projects" (p. 205). The general explanation of this "obvious" fact is that transport is considered as a necessary evil to reach another location and has no utility in itself. Transport would only come about if the excess utility of the other location is higher than the costs of transport, including transport time. Speeding up transport enables to spend less time on transportation and use the saved time for more useful activities. A second explanation why a speed change "obviously" affects travel time rather than distance is the existence of inertia for distance changes; the locations of frequently visited places like home, work, and school are fixed in the short run, making time the only variable that can directly be affected by a speed change.

The notions that travel time has no utility in itself and that persons will use saved travel time for other activities is not supported by research on travelling. First, some travel segments derive a utility from themselves; this is evident for the rather small segment of indirected travelling, like going for a ride or walk. However, Mokhtarian and Salomon (2001) demonstrate that directed travel may generate a positive utility to a significant extent as well. Banister (2011) argues that "travel time can be seen as a social construct, the quality of which should be maximised and more highly valued" (p. 957). Second, studies on travel time spending find that the average time persons spend on travelling is rather constant "despite widely differing transportation infrastructures, geographies, cultures, and per capita income levels" (Schafer, 1998, p.459). This finding suggests that people will not adapt their time allocation when the speed of the transport system changes.

The objections to the focus on travel time incited some authors to propose an alternative method for the assessment of a speed change. Starting from the notion that distance is the mainly influenced variable, the value of a speed increase would regard the benefit of the increasing range of travelling. Metz (2008) and Cervero (2011) assume that a better access is the main benefit of a speed increase. Metz argues that "the bulk of the economic benefit of road schemes and other transport infrastructure investment is associated with making possible additional access to desired destinations" (p. 326).

The idea of improved access raises some objections as well. Whereas the theory that saved travel time is the benefit of a speed increase assumes a constant travel pattern (implying no impact on distance), the theory of increased access assumes a constant land use, that is: no impact on locations of living, jobs, facilities. If a speed change would affect land use, the impact on the access is undefined and an increase will not necessarily improve the access. As we will argue later in the paper, speed changes do affect land use in a way that reduces the initial impact on access.

In the paper, we do not assume beforehand how a speed change works out, but we start with the examination of the (long-term) impacts of a speed change. The outcome enables to identify the benefits of a speed increase. Two research questions are addressed: 1) what are the main impacts and benefits of an increase of travel speed; and 2) is the value of time, as it is generally calculated, a good proxy for the value of speed?

The search for the impacts starts with a discussion on the associations between speed and travel time (Section 2), and between speed and access (Section 3). Section 4 proposes an alternative affected variable: human wealth, and discusses the benefits of a wealth increase. The question whether the value of time can be used as a proxy for the value of speed, assuming that wealth is the most affected variable, is discussed in Section 5. The conclusions are presented in Section 6.

## 2. The association between speed and travel time

A number of studies on time spend on travelling demonstrate that the average travel time per person per day is rather invariant, somewhat more than one hour (Zahavi and Talvitie, 1980; Schafer, 1998). This finding is generally used for town planning and for travel modelling, but it is relevant for the assessment of a speed increase as well. The observation of an invariant travel time suggests that the assumed travel time savings in the traditional assessment are zero. It raises the question whether speed and travel time are associated or not.

We are not aware of any study that directly assesses this question, though a number of studies give the opportunity to examine the association between speed and time. One is the study edited by Szalai (1972), one of the most influential studies on travel time spending. In a number of European and American cities data on average travel time and on the modal-split in commuting to work were collected. The modal-splits vary widely, ranging from walking is the dominant mode (77% in the city of Kragujevac, Yugoslavia/Serbia) to car is dominant (92% in the city of Jackson, USA). We estimated from the modal-split figures the average speed for the residents of each city, assuming that the average speed by mode was equal to that observed in the UK (Department for Transport, 2005). Figure 1 shows the estimated speeds and the observed travel times for the surveyed cities. The speed figures regard only commuting trips, the time figures regard all travel. Unfortunately, the data give not the opportunity to estimate the average speeds for all trips; these will likely differ from the commuting speeds. However, it is a plausible assumption that the speeds for commuting and all travel are strongly correlated and that using speed figures for all trips will produce a similar graph.

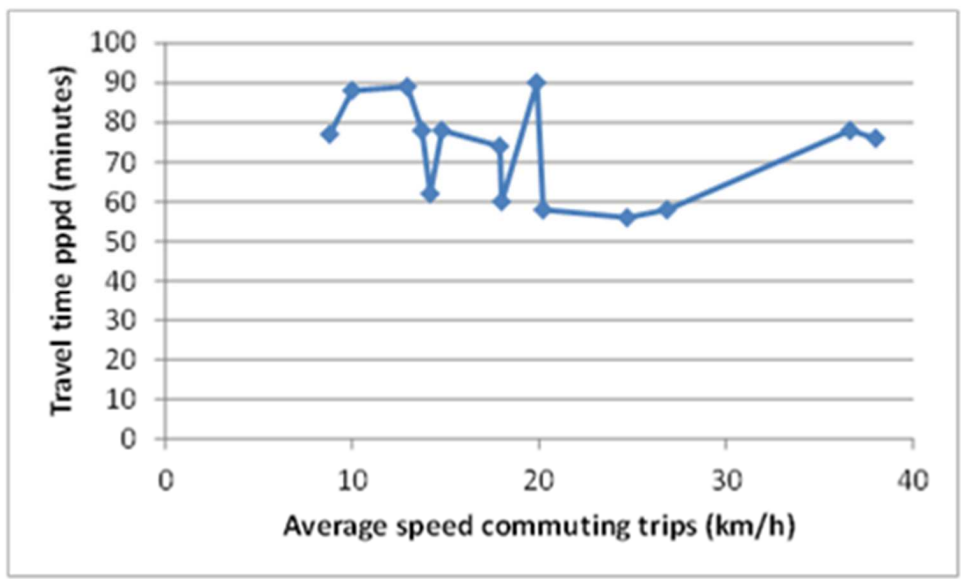

Figure 1 : Speed and travel time for different cities; the 2nd to 4th most right points do not refer to one city but are averages for a number of cities

Estimation of the model 'time = b0 + b1*speed' produces no significant value for b1 (P = 0.43). On the other hand, the association between speed and travel distance is very strong (P = 0.0000008).

In contrast to Szalai who compared travel times in different cities at the same time period, other studies address the development of travel time spending over time in the same spatial context. Zahavi and Ryan (1980) analysed the stability of some travel components in two large US cities in a period with a significant increase of the car travel speed. They identified three groups of travellers: car only users, transit only users, and mixed mode users. Table 1 shows the main findings.

Table 1 : Speed and mobility trends in Washington D.C. (1955-1968) and Twin Cities (1958-1970)

|  | Washington D.C. | | | Twin Cities (Minneapolis-St. Paul) | | |
| --- | --- | --- | --- | --- | --- | --- |
|  | Car only users | Transit only users | Mixed mode users | Car only users | Transit only users | Mixed mode users |
| Increase in: | | | | | | |
| - Speed[1] | 24% | -7% | 15% | 33% | 1% | 25% |
| - Distance[2] | 26% | 6% | 18% | 32% | 10% | 27% |
| - Time[2] | 2% | 13% | 2% | -1% | 10% | -1% |

1: per trip (door-to-door); 2: per traveller

The table shows a strong correlation between speed and distance and no clear correlation between speed and time for the two groups that use the car. However, the picture for the transit only users is quite different. Travel time spending of this (small) group, that did not benefit from the increased car speed, increased. The increase cannot fully be explained by a speed decrease of transit: the speed decreased in just one city and to a smaller extent than the time increase. The authors state that the increase of the car speed was the result of "additional highway capacity and a general reduction in density of development" (p. 20). The reduced density of development is likely the main driving force behind the travel time increase for transit only users.

National travel surveys provide data for analysing mobility trends on a national scale. Figure 2 shows the trends of average travel time, distance, and speed in the UK in a 30-year period with a strong speed increase, from statistics of Department for Transport (2005). The figure illustrates, again, that speed is associated with distance unlike with travel time. Moreover, it suggests that the inertia for adapting the distance play no significant role: no time lag is visible between the speed and distance curves. A possible explanation is that "economic actors are often rational in foreseeing growth in capacity and may often respond prior to its opening" (Noland, 2008).

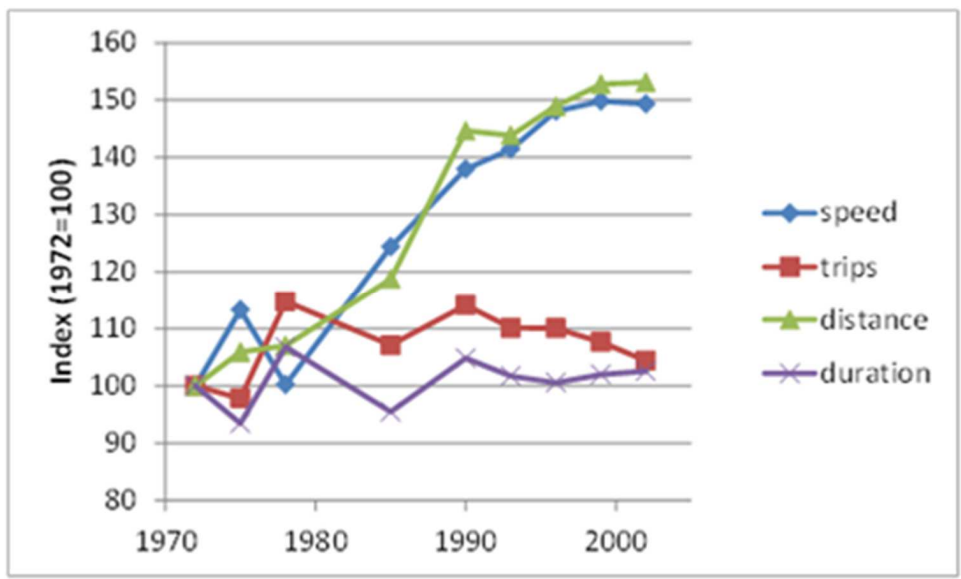

Figure 2 : Trends of mobility indicators for British residents in the period 1972-2002; source: Department for Transport (2005), Table 1.1

Figure 2 includes the trend of the trip rates as well. Just like the travel time, the trip rates are not affected by the speed increase. This is in line with the findings of Hupkes (1982) and most of the findings of Zahavi and Ryan (1980); Zahavi and Ryan observed one exception: a small decrease in the trip rates of transit only users in Washington D.C., the city where the speed of transit decreased somewhat. We conclude that there is strong evidence that the major impact of a speed increase is lengthening of the trips, leaving trip time and trip rates (more or less) unaffected.

The observed stability of travel time is generally indicated as travel time budget. The concept of a budget suggests that people are inclined to spend a fixed part of their time on travelling and will adapt their travel pattern when an external event like a speed increase affects the initial travel time. The stability of travel time would then be explained by compensation behaviour of travellers. If this explanation is correct, the question arises why one group in the before mentioned studies, the transit only users in the study of Zahavi and Ryan, does not (fully) compensate for the initial increase in travel time. Is the response on travel time changes asymmetrical in the sense that travel time reductions are compensated for and travel time increases not? This is not a plausible assumption, considering that an initial travel time increase entails a certain pressure to reduction (assume the extreme case that travelling would take more than 24 hours per day; this is impossible) while a corresponding pressure is not valid for an initial time reduction. Probably, compensation behaviour cannot fully explain the stability of travel time; other factors may play a role as well.

Referring again to the two American cities, the changed density of activities was another factor that affected travel time. For the car only users, this factor was 'helpful' in compensating for the travel time reduction resulting from the speed increase. For the transit only users who were faced with the lower density of activities as well, no speed increase could partly compensate for the increased travel times. Apparently, compensation behaviour as a single factor could not fully neutralize the increased travel times. The generally observed stability of travel time might be explained by both compensation behaviour and the impacts of speed changes on land use. The assumption behind is that speed changes affect the location of activities in a way that partly reduces the initial change in travel time, and that in the case of the American cities the lower density of activities was induced by the car speed increase. The next section includes a discussion on the association between speed and land use that will underpin this assumption.

### 3. The association between speed and access

The finding that speed has no clear influence on travel time implies that a speed increase enlarges the distance travelled. Distance has no clear benefit in itself, as little as speed, but the opportunity to travel longer distances increases the range of travelling. A strand of literature defines improved access as the benefit of a speed increase (Metz, 2008; Cervero, 2011). This raises the question what is the association between speed and access.

For the discussion, we define three access related concepts. The first is 'proximity' which relates to the distance that has to be bridged to reach a certain activity. The second is 'access' which relates to the travel time that has to be spent to reach a certain activity. The third is 'accessibility' which relates to the travel time to a certain geographical location. Then proximity is defined by land use and is independent from speed; access is defined by both land use and speed; and accessibility is defined by speed and is independent from land use[1].

An increase of travel speed implies an improved accessibility and would improve access as well if land use would not be affected by the speed increase. However, if this condition is not satisfied, the impact of a speed increase on access is undefined and the access will not necessarily be improved. This raises the questions: does a speed increase affect land use and if so, how will the land use change turn out?

---

[1] Similar definitions of access and accessibility can be found in the literature. Metz (2008) clarifies his concept of access as "access to desired destinations" (p. 324). And Hansen (1959) uses the concept of accessibility according to our definition when he examines "how accessibility shapes land use" (paper title)

There is strong evidence and a broad agreement that land use is to a large extent shaped by accessibility (Hansen, 1959; Wegener, 1995; Noland, 2008). An improved accessibility, which means a higher speed, raises the land value. When land value increases, value is placed on land that was not accessible before and enlarges the area of land that will be developed. This implies that activity locations will be scattered over a larger area and the distance between the activities will increase. The positive impact of the increased accessibility on access is at least partly undone by a negative impact of a decreased proximity.

Another impact of an improved accessibility is a reshuffling of the visits of locations. This affects the viability of services on certain locations and may induce a changed spatial distribution of the provision of a service. Take as an example shops in the period with a rapid increase of car dependency and related strong speed increase. The higher speed enticed residents of small settlements to shift for their daily shopping from the local shop to the more distant supermarket in the regional city that offers a larger selection at lower prices. Though proximity is an important factor for the food store choice, other factors like quality of the foods, widest selection and best prices are important as well (Handy and Clifton, 2001). As a consequence, local shops might not survive, forcing both car owners and others to travel to the city for shopping. The impact of the speed increase is in this case a concentration of shopping facilities, implying a decrease of proximity. The positive impact on accessibility is at least partly reversed by the negative impact on proximity, and the outcome for access is undefined. For those who did not use a car and did not benefit from the speed increase, the result was an unambiguous decrease of the access.

Both the increased land value and different visiting pattern decrease the proximity of destinations. Then a speed increase has two opposite impacts on access; a direct positive impact and an indirect negative impact. The association between speed and access depends on the relative size of these two impacts.

A valuable study in this respect is Marchetti (1993) who analysed the spatial expansion of cities in relation to technical improvements in the transport system. He observed that ancient cities had never a radius exceeding 2.5 km which fits with a walking speed of 5 km/h and a time budget of 1 hour. When introducing mechanical transportation with higher speeds, cities started to grow, and the expansion was closely associated with the stepwise introduction of faster travel modes. This finding suggests a tendency to a constant access. The expansion implied a lower proximity of the city centre and its activities, but the concurrent increase of both the city radius and the transport speed left the access more or less unaffected. The two opposite impacts on the access would be more or less balanced, implying no clear association between speed and access.

The impact of a speed increase on proximity suggests that the initial proximity was to some extent pinching. The speed increase relieves the tightness of proximity and creates the opportunity for land use changes that reduce proximity. The examples so far concerned shops and other amenities. Would a speed increase have a similar impact on the proximity of other types of locations, like eligible jobs or homes of family or friends? For a good valuation of speed, research on the association between speed and proximity of different types of locations is necessary. If there is an association, a speed increase will likely relieve the tightness of proximity and then counteract the impact on accessibility. This would mean that some disadvantages are connected with the need for a certain proximity and that reducing this need generates some benefits. These benefits are –apart from a possible limited increase of access– the true benefits of a speed increase. These are discussed in the next section.

### 4. The benefit of a speed increase

If a speed increase does not reduce travel time and has no clear impact on access, are there alternative benefits and if so, what is the nature of the benefits? For the discussion one should identify what actually the impacts are of an increase of the transport speed. Taking the example of shops that cannot survive in small settlements, the impact is a concentration of shopping facilities in regional cities. These provide a larger selection at lower prices than the local shops, and the concentration brings about scale economies that affords to widen the selection and lower the prices further. The reduction in the density of development of cities implies more room for living and other activities. In addition to the impacts for person travel, a speed increase in freight transport gives the opportunity to transport perishables to more distant locations so widening the selection for consumers; it also facilitates concentration of production activities which generates economies of scale. These impacts can be summarised as an increase in human wealth. The benefits of a speed increase can then be translated into the benefits of a wealth increase.

We can add here that in some cases of improved access, the nature of the benefits is a wealth increase as well. Assume the case that local shops manage to survive despite a fall in business. Then the proximity is unchanged and the access is improved. The nature of the improved access is more people having access to a larger selection of products at lower prices, which is an increase in wealth. To make it more generally, the benefit of a higher access at given travel times is a higher utility of the visited destinations. The nature of the excess utility will be in many cases (though not all cases) an increased wealth. Next, we hypothesize that an increase in wealth or prosperity is the major impact of a speed increase and discuss these benefits.

One can argue that the size of the benefit of an increase in wealth or prosperity is associated with the prosperity level of a society. Following the law of diminishing marginal utility, the benefits of a marginal wealth increase will be larger in poor countries than in wealthy countries. Studies in the field of happiness science support this argument. Veenhoven (1991) and Lane (2000) find a clear decreasing positive correlation between happiness and prosperity when comparing different countries. There is even evidence that in wealthy countries an increase in income does not affect happiness (Easterlin, 1995) or has a reverse effect; in some of the most wealthy countries a slightly decreasing trend is observed for happiness despite a continuing increase in the national product (Lane, 2000, and Layard et al, 2010, for the United States; Ferrer-i-Carbonell, 2005, for Western Germany). Though they observe that inhabitants with higher incomes are happier, an increase of the overall income has no impact or a slightly negative impact on the happiness of the whole population. The explanation is that in these countries people are more concerned about their relative income than about their absolute income. When in a wealthy country the income of everyone increases at the same rate, the negative effect of the increase of the income of the social reference group exceeds the positive effect of the increase of the own income. The explanation why an increase of the absolute income for everyone (no change in the relative income) has no marked positive impact is the adaptation of aspirations. The presumption is, that "once basic needs are met, aspirations rise as quickly as incomes, and individuals care as much about relative differences with their peers as they do about absolute gains" (Graham et al, 2010, p. 248). Vendrik and Woltjer (2007) demonstrate that the function describing the dependence of life satisfaction on the relative income is concave, both for positive and negative relative incomes. This implies that a change in the income *distribution* can have a significant effect on happiness; levelling off increases happiness, enlarging the income differences would have the opposite effect.

Frey (2008) argues that the concept of happiness can be used as a measurable proxy for the abstract concept of utility, considering that happiness is for many people an ultimate goal. It is true that there are some drawbacks. First, the concept of happiness is not clearly defined. Diener et al (2010) explain that measures of happiness includes the components judgement and affect and can have various emergences depending on the degree each of these components is included. Second, happiness is not the only goal that matters; examples of other goals are responsibility and health. Still, "happiness is undoubtedly an overriding goal in most people's lives" (Frey, 2008, p.5). Variables that affect happiness can be assumed to affect utility in a similar way.

Assuming that a) the predominant effect of a speed increase is growth of human wealth, and b) the benefits of a wealth increase can be properly measured by the impact on happiness, the benefits of a speed increase will generally be larger in poor societies than in wealthy societies. In the latter, the benefits could be marginal or even absent.

## 5. The value of travel speed versus the value of travel time

This section discusses how the value of time that traditionally is used for assessing the benefits of speed changes relates to the value of speed. Is the value of time a good proxy for the value of speed? Having in mind the relation between the value of travel speed and prosperity, we start the discussion with an examination of the association between the value of travel time and prosperity.

Initially, the value of saved travel time was assumed to be equal to the earnings that would have been received if this time was used for economic production. The value of time then equals the individual wage rate (Jara-Diaz, 2007). However, it was noticed that saved time is not always fully employed for productive activities; part of the time saved may be used for leisure activities. An alternative to the wage rate is the willingness to pay for saving a certain time period. This has a more general value than the wage rate and is the most commonly used measure for assessing the benefits of time savings. The willingness to pay is a subjective factor that varies for different individuals and different situations. It is related to the characteristics of travellers and trips. Particularly high-income people and business travellers generally have a high willingness to pay.

The value of saved travel time proves to be positively correlated with prosperity. Gunn (2007) mentions an income elasticity of 0.5, implying that a 10% rise in income would increase the value of time by 5%. Mackie et al (2003) recommend an elasticity of 1.0 for business travel and an elasticity of 0.8 for non-working purposes. The Department for Transport (2015) confirms the 1.0 for business travel and suggests that the elasticity for non-working purposes could be 1.0 as well but stresses the uncertainty. The consequence is a long term growth of values of time in the range of 1.5-2% per annum (Rus and Nash, 2007). The association between prosperity and value of time can be explained simply. If the value of time is based on wages, the association is obvious; prosperity is directly related to income. If the value of time is based on willingness to pay, the argument is simple as well: when prosperity increases, people have a higher ability to satisfy their needs, leaving less important needs unsatisfied. Spending of additional money on the alternatives for travel time saving is then less beneficial, implying a larger willingness to pay for saving travel time.

The observed positive correlation between the value of time and prosperity is in the opposite direction of the negative correlation between the value of speed and prosperity that was argued in the preceding section. The different correlations are shown in Figure 3. The value of speed is indicated by a convex declining curve, corresponding to the derivative of a utility function that increases with prosperity to a decreasing extent. The assumption behind the value of time curve is a constant elasticity with respect to prosperity.

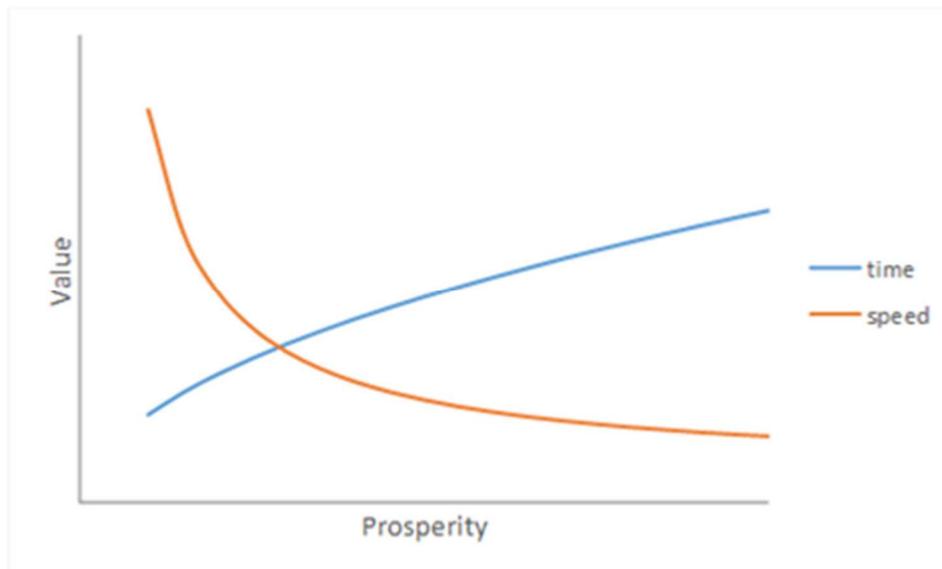

Figure 3 : The values of speed and time related to prosperity

We conclude that the value of time and the value of speed are basically different and therefore the value of time cannot be used as a proxy for the value of speed.

One might wonder about this conclusion. The willingness to pay for a travel time reduction is essentially not different from the willingness to pay for a speed increase. Even if we observe that a speed increase does not affect travel time, one can adopt the argument of Goodwin (1981) that the rationale of valuing travel time "is based entirely on the assumption that, when saved, it will be used for some unstated alternative purpose, valued because it brings some utility to the traveller or to somebody else, now or in the future… If time is saved from one journey and the traveller chooses to spend it on another journey exactly the same logic applies; the value of time saved is now being used as a proxy for the utility of a wider choice of destinations" (pp. 99-100). Considering the fact that because of the impact on land use no "wider choice of destinations" might exist, the argument is still valid for any alternative benefit of a speed increase. Van Wee and Rietveld (2008) argue that even the alternative benefit will likely be larger than the benefit of the travel time savings because otherwise people would reduce travel time rather than choose additional travel. Though one could argue that the willingness to pay will be a less accurate measure for the value of speed when the nature of the benefits is less imaginable for the payer, valuing speed based on the willingness to pay would not be basically wrong.

For the explanation we will discuss two serious shortcomings of using the willingness to pay, or, more generally, observed choice behaviour, as an indication of utility. One shortcoming regards misprediction of the personal benefits that will be gained from a decision (Frey, 2008). People make systematic errors in the prediction and their choices will not maximize their personal utility. An important reason for the incorrect prediction is a general underestimation of the ability to adapt to a changing situation. The adaptation mitigates the impact of an event (either a negative or positive impact) and when the adaptation is underestimated, the predicted impact will overstate the actual impact. This argument states that the willingness to pay is a poor indication of utility and will generally overstate it, but it does not explain the opposite direction of the two curves in Figure 3.

The second shortcoming regards neglect of the general, social benefits. The willingness to pay does not take into account the impact on the social utility. One of the findings in happiness studies is that happiness is associated with relative income. An income or wealth increase of some persons may make others unhappier because their relative position becomes worse. This negative impact of a wealth increase is neglected by the willingness to pay; this reflects only the individual preferences that are directed at satisfying the personal needs. Take as an example two neighbours who both prefer to own the most prestigious car of the street. If the one who owns the simplest car buys a new one that is more prestigious than that of his neighbour, he makes his neighbour unhappy. The intervention creates a disutility that could equal the excess utility of the new car for the buyer. This shortcoming gives an explanation for the two opposite curves.

The social comparison can affect the personal utility in a negative way as well. The comparison may create a preference for a certain status that is defined by the general position of people belonging to the same social group. The aim for a status can induce someone to make choices that lower the own utility. Consider a commuter who is happy with the daily congestion because it lengthens the quiet period between the busyness at home and workplace. Still, if a pay lane would be introduced enabling to drive uncongested, he might pay for faster driving just because it is not done for his status to spend time in congestion if there is a faster alternative.

6. Conclusions

Generally, the value of speed is assessed as the value of travel time that would be saved by a speed increase and reallocated to more useful spending. This assumes an association between speed and travel time. However, research on travel behaviour gives no evidence about this association. Based on this finding, improved access is proposed as an alternative benefit of a speed increase. This assumes an association between speed and access. This association is unclear and likely weak; the initial impact on access is at least partly undone by a reverse impact of the speed increase on proximity. We assume that a wealth increase is the major benefit of a speed increase. The value of speed based on this assumption is basically different from the value of time that assumes time savings. The wealth assumption implies that the value of speed is negatively correlated with prosperity, while the value of time proves to be positively correlated. As a consequence, the value of speed might be underrated by the value of time in poor societies and overstated in wealthy societies. At which prosperity level the underrating reverses in overstating can be subject for further research. We assume that in the developed countries the value of speed is largely overstated, considering the finding that in these countries a general wealth increase has hardly any impact on happiness; happiness can be used as a proxy for utility.

The discussion in the paper is limited to the internal benefits of a speed increase. There are a number of external effects as well, like impacts on traffic safety, emissions of greenhouse gasses or pollutants, and noise nuisance. If the speed increase would affect the income distribution, this would be an external effect as well. Since the external effects are not or weakly associated with prosperity, they will become relatively more important when prosperity increases. In the wealthy developed countries the external effects may be the dominating effects of speed changes.

The discussion is on a conceptual level. The paper does not provide a definite method for the value of speed assessment; developing such a method could be a subsequent step. Additionally, more research on the impacts of speed changes is recommended. Is it true, that the benefit of a speed increase is predominantly an increase in human wealth, the basic assumption in the paper? To which extent is the initial impact of a speed change on access undone by a change in proximity? And to which extent is the observed invariance in travel time spending the result from compensation behaviour of travellers or from changes in proximity?